\documentclass[prl,twocolumn]{revtex4}
\usepackage{graphicx}
\begin{document}
%\draft
%\preprint{}
\title{Magnetic imaging of Pearl vortices}
\author{F. Tafuri$^{1,2}$, J.R. Kirtley$^2$,
P.G. Medaglia$^3$, P. Orgiani$^3$ and G. Balestrino$^3$\\}

\affiliation{$^{1}$INFM, Dip. Ingegneria dell'Informazione, Seconda
Universit\`{a} di Napoli, Aversa (CE), Italy}
\affiliation{$^{2}$IBM Watson Research Center, Route 134 Yorktown Heights, NY, USA}
\affiliation{$^{3}$INFM-Coherentia, Dip. Ingegneria Meccanica, Universit\`{a} di Roma Tor Vergata, Roma, Italy}

\date{\today}

\begin{abstract}
We have used scanning SQUID magnetometry to image vortices
in ultra-thin
[Ba$_{0.9}$Nd$_{0.1}$CuO$_{2+x}$]$_m$/[CaCuO$_2$]$_n$ (CBCO)
high temperature superconductor
samples, with as few as three superconducting CuO$_2$ planes.
The Pearl lengths ($\Lambda = 2\lambda_{L}^2/d$, $\lambda_{L}$ the
London penetration depth,
$d$ the superconducting film thickness) in these samples, as
determined by fits to the vortex images, agree with those
by local susceptibility measurements, and
can be as long as 1mm. The in-plane
penetration depths $\lambda_{ab}$ inferred from the Pearl lengths
are longer than many bulk cuprates with comparable critical temperatures.
We speculate on the causes of the long penetration depths,
and on the possibility of
exploiting the unique properties of these superconductors for basic experiments.
\end{abstract}

\maketitle

%\begin{multicols}{2}
%\narrowtext
%\pacs{73.40.Gk,73.40.Rw,74.50.+r,74.72.-h}

%\pagebreak
%\narrowtext

%\section{Introduction}

Vortices play a central role in many aspects of superconductivity.
Not only do the dynamics of vortices determine many of the transport
properties of type II superconductors,
especially the high critical temperature cuprates \cite{blatter},
but vortices are also of more general interest, since as topological
defects they are
of great relevance, for instance, to phase transitions \cite{volovik,mermin}.
The formation of topological defects in phase transitions has
even stimulated some analogies between cosmology, gauge theories
and condensed matter physics \cite{zurek,kibble}.
Vortices in bulk type II superconductors were first predicted by
Abrikosov in 1957 \cite{abrikosov}, and have since been imaged by
many different experimental techniques \cite{bending}.
Vortices in thin superconductors
($d << \lambda_L$, where $d$ is the superconducting film thicknes and $\lambda_{L}$ the
London penetration depth respectively)
were first described by Pearl \cite{pearlapl}
(hence ``Pearl" vortices).
Pearl vortices have several interesting
attributes. The field strengths $h_z$ perpendicular to the films diverge as
$1/r$ at distances $r << \Lambda$ in Pearl vortices, whereas in Abrikosov vortices the
fields diverge as $\ln(r/\lambda_L)$ \cite{kogmints}. Since in the Pearl vortex much of the
vortex energy is associated with the fields outside of the superconductor,
the interaction
potential $V_{int}(r)$ between Pearl vortices has a long range component
$V_{int} \sim \Lambda/r$ for $r >> \Lambda$ \cite{pearlapl}, unlike Abrikosov vortices,
which have only short range interactions.
The interaction between Pearl vortices
$V_{int} \sim \ln(\Lambda/r)$ for $r << \Lambda$ leads to a
Berezinskii-Kosterlitz-Thouless (BKT) transition which is cut off due to screening
on a scale $\Lambda$ \cite{blatter}. The logarithmic interaction makes this system
very similar to a Coulomb gas and ideal to study screening effects and renormalization in BKT transitions \cite{schei}.
%{\it Due to long-range intervortices interaction near the surface,
%the vortices do not assume a particular lattice order near the
%superconductor surface \cite{pearlapl}.  Finally the presence of a surface modifies the bulk scenario in three ways:
%via the appearance of stray magnetic fields, the lack of protecting superconducting layers,
%and the enhanced vortex fluctuations.}
While superconducting vortices in films with thickness $d$ comparable to
the London penetration depth $\lambda_{L}$ have been imaged using many
techniques, to our knowledge the present work is the first to directly demonstrate
experimentally the existence of Pearl
vortices for $d << \Lambda$, and
is also the first to use
scanning susceptibility measurements to determine
penetration depths in superconductors.

In the present work, two different types of [Ba$_{0.9}$Nd$_{0.1}$CuO$_{2+x}$]$_m$/[CaCuO$_2$]$_n$ (CBCO) structures
were grown: a) the ultrathin
[Ba$_{0.9}$Nd$_{0.1}$CuO$_{2+x}$]$_{M}$/[CaCuO$_{2}$]$_{N}$/[Ba$_{0.9}$Nd$_{0.1}$CuO$_{2+x}$]$_{M}$
($M/N/M$) structure which consists of only one superconducting infinite layer (IL) block ($N$
CaCuO$_{2}$ unit cells), sandwiched between two charge reservoir (CR) blocks ($M$\ Ba-based
unit cells) and the similar $M/N/M/N/M$ structure (M=5 and N=2); b) the thick
[(Ba$_{0.9}$Nd$_{0.1}$CuO$_{2+x}$)$_{m}$/(CaCuO$_{2}$)$_{n}$]$_{S}$
($m\times n$ superlattice)\ structure which consists of $S$ sequences
(with $S\geq$ 15) of the
(Ba$_{0.9}$Nd$_{0.1}$CuO$_{2+x}$)$_{m}$/(CaCuO$_{2}$)$_{n}$\ supercells
composed of $m$ Ba-based and $n$ Ca-based unit cells. All the samples were
grown on (001)\ SrTiO$_{3}$ substrates, with nominally zero miscut angle, by
Pulsed Laser Deposition (PLD), using a focussed KrF excimer pulsed laser source ($%
\lambda$ =248nm) with
energy areal density on the target surface of 7 J/cm$^{2}$ in a spot size of 2 mm$^{2}$.
Two sintered
powder targets, with a nominal composition of (Ba$_{0.9}$Nd$_{0.1}$)CuO$_{2}$
and CaCuO$_{2}$, mounted on a multitarget system, were used.
The substitution of 10\% of the Ba atoms with trivalent Nd cations, even if not
strictly necessary for superconductivity \cite{balestrino0,balestrino apl},
helped to find the right growth
conditions
%for good quality artificial structures
by slightly decreasing the uncompensation
of the electrical charge in the CR block.
The growth temperature was about 640$^\circ$
C and the molecular oxygen pressure was $\approx$ 1mbar. At the end of the
deposition procedure, an amorphous protecting layer of electrically
insulating CaCuO$_{2}$ was deposited on top of the film at a
temperature lower than 100$^\circ$
C.

The SQUID microscope measurements were made
at 4.2 K with the sample cooled
and imaged in fields of a few mG, sufficient to trap several vortices in a 200$\mu$m $\times$
200 $\mu$m scan area. Two types of SQUID sensors were used: 1)
magnetometers \cite{ssmapl} with either square pickup loops 7.5$\mu$m on a side, or
octagonal pickup loops 4$\mu$m in diameter; and 2) SQUID susceptometers \cite{suscapl}
with a single turn field coil 20$\mu$m in diameter, with a square pickup loop 8$\mu$m across
(see Fig. \ref{fig:cbcofig2}a).

%\section{Scanning SQUID Microscopy results and Discussion}

\begin{figure}
\includegraphics[width=3.4in]{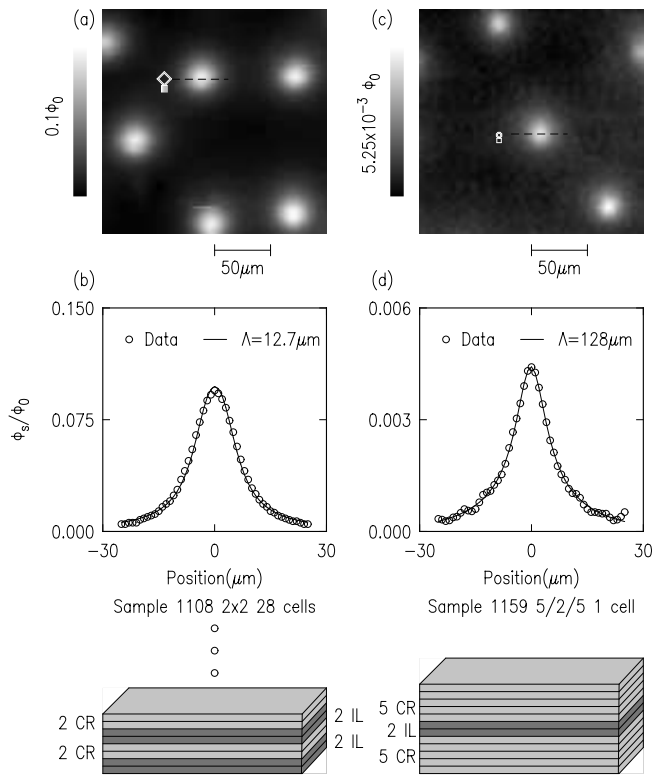}
%\includegraphics[width=3.0in]{cbcofig1.eps}
%\vspace{0.1in}
\caption{SQUID microscope image and cross-sectional data (along the
positions indicated by the dashed lines in (a,c))
of vortices trapped in two CBCO samples.
%1108 (2$\times$2, 28 cells) (a,b) and 1159 (5/2/5, 1 cell) (c,d).
The SQUID pickup
loops were a square 7.5$\mu$m on a side (a,b) and an octagon 4$\mu$m on a side (c,d).
The open symbols in (b,d) are the cross-sectional data; the solid lines in (b,d)
are fits to Eq. (\ref{eq:thinvort}).
Scaled
schematics of the pickup loops used appear in (a,c).}
\label{fig:cbcofig1}
\end{figure}

We have performed scanning SQUID microscopy (SSM) on various $5/2/5$ monolayers and
CBCO-$m \times n$
samples, and as a function of the number of the CuO$_2$ planes. In all systems
we have clearly observed Pearl vortices.
This provides evidence of superconductivity
complementary to
traditional transport measurements \cite{bales1}
for the thinnest films.
We show in Fig. 1 SSM images of vortices trapped in two typical samples.
The thin-film limit for the
two-dimensional Fourier transform of the $z$-component of
the field from an isolated vortex trapped
in a thin film is given by \cite{pearlapl,kogpearl}:
\begin{equation}
h_z(k,z)=\frac{\phi_0 e^{-k z}}{1+k \Lambda},
\label{eq:thinvort}
\end{equation}
where $z$ is the height above the film, $k=\sqrt{k_x^2+k_y^2}$,
$\Lambda=2\lambda_{ab}^2/d$ is the Pearl penetration length,
$\lambda_{ab}$ is the in-plane penetration depth, $d$ is
the film thickness and
$\phi_0 = hc/2e$. We fit the data in Fig.
\ref{fig:cbcofig1} by inverting
Eq. (\ref{eq:thinvort}) to find $h(x,y,z)$, integrating the result
over the known pickup loop geometry, and using
$\Lambda$ as the fitting parameter. The height
$z$ was determined by fitting images of vortices in Nb with the same
SQUID magnetometer, assuming an isotropic
low temperature Nb penetration depth
of $\lambda$=0.05$\mu$m \cite{nblambda}.
We note that although the peak SQUID flux $\phi_s$ depends
strongly on $\Lambda$, the full-width at half-maximum of the
vortex images is relatively independent of $\Lambda$ for such
thin films. For comparison, Abrikosov vortices typically
couple about 0.5$\phi_0$ of flux into the SQUID sensor in
this geometry, and are resolution limited.

The results for the Pearl lengths $\Lambda$
from such fits to images of vortices
for a number of superlattice samples are summarized in Table I. The values
for the film thicknesses $d$ and IL layer thickness $d_{IL}$
were obtained by assuming a thickness/layer
of 3.2\AA \, for the CaCuO$_2$ (IL) layers, and 4.4\AA \, for the BaCuO$_x$ (CR) layers.

\begin{table}[b]
\caption{Pearl lengths $\Lambda$ of various CBCO samples. T$_c$ is measured by
standard four-probe techniques and refers to zero resistance. We estimate
uncertainties in $\Lambda$ of $\pm$20\% and of T$_c$ of $\pm$0.25K.}
\label{tab:tetgroup}
%\begin{tabular} {|c|c|c|c|c|p{60pt}}
\begin{tabular} {|c|c|c|c|c|c|p{43pt}|}
\hline
Sample            &Type                       &Cells     &d(\AA) &d$_{IL}$(\AA) &T$_c$(K) &$\Lambda$ ($\mu$m)         \\ \hline
%1179              &4$\times$2                 &20        &480    &480             &30        &59                   \\ \hline
1159              &5/2/5                      &1         &50     &6.4              &30        &128                  \\ \hline
1151              &5/2/5                      &1         &50     &6.4              &35        &205                  \\ \hline
1988              &5/2/5/2/5                  &1         &79     &12.8             &50        &292                  \\ \hline
1984              &5/2/5/2/5                  &1         &79     &12.8             &50        &490                  \\ \hline
1987              &5/2/5/2/5                  &1         &79     &12.8             &50        &810                  \\ \hline
1985              &2$\times$2                 &12        &182    &76.8             &78        &25                   \\ \hline
1201              &2$\times$2                 &20        &304    &128              &65        &13.6                 \\ \hline
1108              &2$\times$2                 &28        &426    &179.2            &70        &12.7                 \\ \hline
1106              &2$\times$2                 &28        &426    &179.2            &75        &9.1                  \\ \hline
1171              &5$\times$2                 &15        &426    &96               &60        &14.2                 \\ \hline
%\hline
% &$\lambda_{ab}$($\mu$m)
%  &1.19
%  &0.56
%  &0.72
%  &1.08
%  &1.40
%  &1.80
%  &0.48
%  &0.46
%  &0.53
%  &0.44
%  &0.55
\end{tabular}
\end{table}

It is remarkable that Pearl vortices can be observed in
films with Pearl lengths up to a millimeter long:
An Abrikosov vortex in a
superconductor with penetration depth $\lambda_{ab}$=128$\mu$m
(Fig. \ref{fig:cbcofig1}(d))
would couple a peak flux of about 1$\times$10$^{-3}\phi_0$
into the SQUID sensor, with a peak width of about 250$\mu$m,
making imaging extremely difficult. However, although
the peak flux from the Pearl vortex in Fig. \ref{fig:cbcofig1}(d)
is relatively small ($\sim 4.5\times10^{-3}\phi_0$),
the strongly
diverging fields in Pearl vortices ($h_z \sim 1/r$, $r$ the in-plane radius
from the vortex center) give
sharp peaks in the scanning SQUID image.
The resulting
strong contrast makes
it feasible to determine the Pearl
length.
A consistency check on the Pearl lengths so determined can be obtained by
making scanning susceptometer measurements \cite{obukhov}.
In the measurements illustrated by
Figure \ref{fig:cbcofig2} the sample is driven down
until it comes into contact with a corner of the SQUID susceptometer,
which is mounted on a
flexible brass cantilever.
Contact occurs at a spacing between the SQUID pickup loop and the
sample surface of about 5.0$\mu$m, as determined by fits of
similar data using a Nb sample, assuming a Nb penetration
depth of 0.05 $\mu$m.
%The mutual inductance between the field
%coil and the pickup loop changes appreciably
%when the pickup loop to sample spacing  becomes comparable
%to the diameter of the field coil.
In principle, the mutual
inductance should saturate when
the SQUID substrate contacts the sample.
Experimentally there continues to be some change,
presumably because the tilt angle
between the substrate and the sample decreases.

\begin{figure}
\includegraphics[width=3.5in]{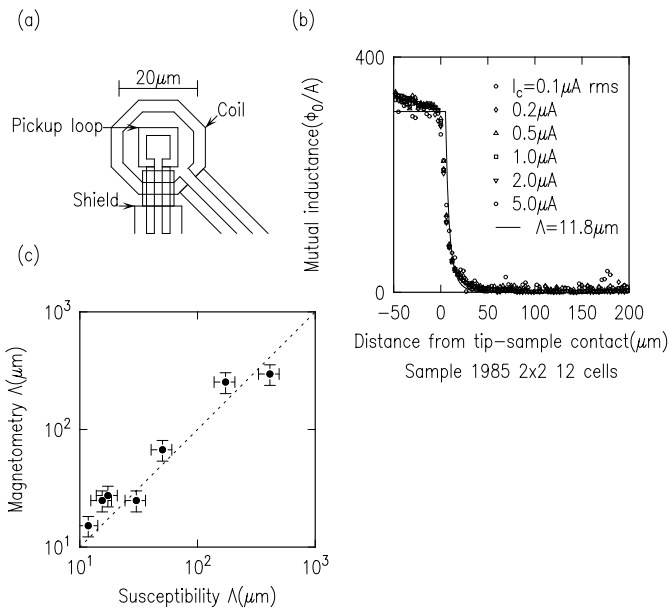}
\vspace{0.1in}
\caption{(a) Geometry of
%concentric, co-planar field coil and pickup loop in the integrated
the SQUID susceptometer used. (b) Mutual inductance between the field coil and the pickup loop,
as a function of the spacing $z$ between the SQUID substrate and sample. The symbols are
data, taken with various alternating currents through the field coil.
The solid line is modelling using Eq. (\ref{eq:kogsusc}), with
$\Lambda$=11.8$\mu$m.
(b) Comparison of the Pearl length $\Lambda$ for a number of CBCO samples
using fitting of SQUID magnetometry images of vortices (vertical axis) vs susceptibility
measurements (horizontal axis).}
\label{fig:cbcofig2}

\end{figure}

The 2-D Fourier
transform of the $z$-component of the field in the
pickup loop, with a current $I$ in a circular ring
of radius $R$ oriented parallel to,
and a height $z$ above a sample,
is given by \cite{kogsusc}
\begin{equation}
h_z(k)=\frac{-4 \pi^2 I R}{c} J_1(kR) (1 - \frac{e^{-2kz}}{1+k \Lambda}),
\label{eq:kogsusc}
\end{equation}
where $J_1(kR)$ is
a Bessel function of the first kind. The solid line in Fig. \ref{fig:cbcofig2}(b)
is obtained by numerically integrating the 2-D Fourier transform of
Eq. (\ref{eq:kogsusc}) over the area of the pickup loop, for various
values of $z$, and fit to the data by varying $\Lambda$.
Figure \ref{fig:cbcofig2}(c) compares the values obtained for the Pearl lengths
for a number of the CBCO samples using magnetometry and
susceptometry methods. The two methods agree within experimental error over
the range of Pearl lengths present. Since the fitting to the Pearl vortex
images was done assuming each vortex has $\phi_0$ of total flux
threading through it, rather than a fractional value \cite{mintsfrac},
this agreement, especially in the $5/2/5/2/5$ sample
means that the superconducting layers are sufficiently strongly
Josephson-coupled when separated
by a CR layer made of five BaCuO$_x$ unit cells to make it energetically favorable for the vortex
flux to thread vertically through the superconducting layers, as opposed
to escaping between the layers.
%This represents an intriguing topic for further studies involving also topological and
%phase transtion issues.

%\begin{figure}
%\includegraphics[width=2.5in]{lpvslp.ps}
%\vspace{0.1in}
%\caption{Comparison of the Pearl length $\Lambda$ for a number of CBCO samples
%using fitting of SQUID magnetometry images of vortices (vertical axis) vs susceptibility
%measurements (horizontal axis).}
%\label{fig:lpvslp}
%\end{figure}

\begin{figure}
\includegraphics[width=3.0in]{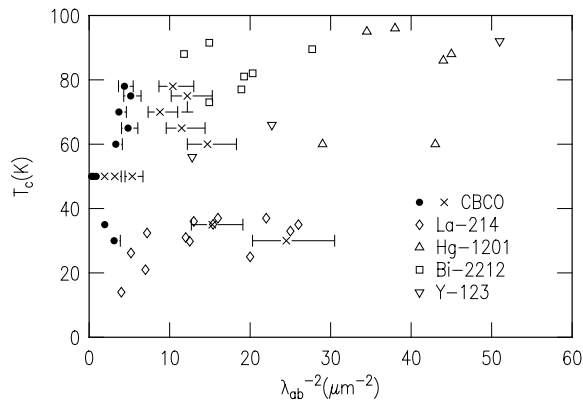}
\vspace{0.1in}
\caption{
Values of T$_c$ vs $\lambda_{ab}^{-2}$.
The solid dots are the results for CBCO, using
$\lambda_{ab,h}=\sqrt{d\Lambda/2}$ (homogeneous distribution
of superfluid density). The crosses are $\lambda_{ab,IL} =\sqrt{d_{IL}\Lambda/2}$
(superfluid density localized in the
IL layers). The open symbols
are recent results
for a number of bulk hole-doped cuprates: La$_{2-x}$Sr$_x$CuO$_{4}$
(La-214) \cite{la214}; HgBa$_2$CuO$_{4+\delta}$ (Hg-1201);
Bi$_2$Sr$_2$Ca$_{1-x}$Y$_x$Cu$_2$O$_{8+\delta}$ (Bi-2212) \cite{hg1201};
YBa$_2$Cu$_3$O$_{7-\delta}$ (Y-123) \cite{y123}.}
\label{fig:uemura}
\end{figure}

As expected, the Pearl lengths are longest for the thinnest
CBCO films.
Fig. \ref{fig:uemura}
shows that the CBCO penetration depths
$\lambda_{ab,h}=\sqrt{d\Lambda/2}$ obtained
assuming a homogeneous film (solid circles)
are longer than
for a number of hole-doped cuprates with comparable
critical temperatures \cite{la214,hg1201,y123}.
For example,
optimally doped YBa$_{2}$Cu$_{3}$O$_{7-\delta }$ (Y-123), with a T$_c$ of 92K, has $\lambda_{ab} \sim$ 0.15$\mu$m
\cite{bonn}. The highest T$_c$ CBCO sample (sample 1985, T$_c$=78K) has
$\lambda_{ab,h} = 0.48 \mu$m.
Our samples
span a wide range of Pearl lengths and sheet resistances per square.
Detailed measurements of the latter are given elsewhere
\cite{balestrino0,bales1}.
The 2$\times$2 superlattices have
resistance per square values a factor of 10 lower than the metal-insulator
limit in the $2\times n$ superlattice series\cite{balestrino0}.
Since the mean free
path in the high resistivity (but metallic, $n \sim 11$ or 5/2/5/2/5)
films can be no shorter than the
width of a CuO$_2$ unit cell ($\sim$4\AA), and since the normal state
carrier sheet densities and effective masses
should be similar within this series,
this implies that the mean free paths in the $2\times2$ superlattices
must be at least a few times larger than
the in-plane coherence length ($\xi \sim$20\AA),
and that the Pippard correction for the
effect of a finite mean free path $l$,
$\lambda_{eff} = \lambda_L(1+\xi/l)^{1/2}$, cannot be large.
The London approximation ($\lambda_L^2=m^*c^2/4\pi n_s e^2$,
\cite{degennes},
where $m^*$ is the effective mass of the charge
carriers) may therefore be
reasonable to evaluate the superfluid density for this type of structure.
If we use the standard London expression
and a reasonable value of
m$^*$= 5m$_e$ \cite{kresin}, we obtain $n_s$=6.29$\times$10$^{21}$ cm$^{-3}$ for
optimally doped Y-123, as compared with $n_s$=6.28$\times$10$^{20}$ cm$^{-3}$
for the highest T$_c$ CBCO sample (1985). The corresponding areal superfluid densities
per plane $n_p=n_sd/N_p $ are 3.67$\times$10$^{14}$ cm$^{-2}$ for optimally
doped Y-123 and 3.2$\times$10$^{13}$ cm$^{-2}$ for CBCO sample 1985
($N_p$ is the number of superconducting CuO$_2$ planes).
The superfluid densities for the CBCO samples are about a factor of
10 lower than for Y-123, although they have comparable T$_c$'s,

It has been proposed that the superfluid screening in films could
be supressed by proximity to the metal-insulator transition \cite{mitransition}
or quantum fluctuations \cite{quantumfluct}. However the 2$\times$2 superlattices
have normal state resistances 10 times smaller than the metal-insulator
critical resistance of $\sim$26kOhm \cite{balestrino0}.
%Indeed,
%these films have high normal state resistances of
%1-2 kOhms/square - not far from the metal-insulator transition \cite{balestrino0}.
It appears
that the penetration depths in these films are significantly larger than
bulk cuprates with comparable T$_c$'s.
This may mean that these compounds are more efficient
at producing high T$_C$'s from a given superfluid
density \cite{uemura2}.

A clue to how this could come about comes from
considering the layered structure of these films.
If instead of assuming that the superfluid densities
are homogeneously distributed, we assume instead
that all of the superfluid density
is localized in the IL layers, then
$\lambda_{ab,IL}=\sqrt{d_{IL}\Lambda/2}$.
In this case the calculated penetration depths
(the crosses in Fig. \ref{fig:uemura})
become comparable to the longest penetration depths
reported for some cuprates: Although the average superfluid
density in these films is low, the density in the IL
layers might be higher, possibly promoting superconductivity
at high temperatures.

We also note that the areal superfluid densities
are about 2$\times$10$^{14}$ cm$^{-2}$ for the
5/2/5 structures, making
them ideal candidates for field effect
experiments.
The height of the surface barrier $E_o$ to formation
of vortices
is one of the crucial parameters to observe vortex quantum tunneling (VQT) \cite{vortextunneling}.
$E_o$ is proportional to $\phi_0^2/(8 \pi^2 \Lambda)$
and therefore inversely proportional to $\Lambda$: the larger the Pearl length,
the lower the barrier height.

%\section{Conclusions}

In conclusion, we have investigated vortex matter  in
ultrathin [Ba$_{0.9}$Nd$_{0.1}$CuO$_{2+x}$]$_m$/[CaCuO$_2$]$_n$ systems
using scanning SQUID magnetometry and
susceptometry.  We have given the first experimental
evidence for Pearl vortices in the regime $d << \lambda_L$.
This can be considered the closest attempt yet
to investigate vortices in 2-dimensional systems (vortices of zero length).
This experiment proves that extreme regimes (ultrathin films) are
experimentally accessible through SSM and opens up several prospects
of broad interest, especially if we consider that these
topological defects may have analogies in other fields of physics.
These measurements
identify systems with very long penetration depths and relatively high T$_c$.
This represents a further step to experimentally isolate the properties
important for superconductivity in high-T$_c$ compounds. Finally,
these systems potentially represent ideal systems to test novel theories
and concepts for devices (VQT and field effect experiments).

This work has been partially supported by the ESF projects ``Pi-Shift" and ``VORTEX".
The authors would like to thank G. Blatter, J. Guikema, V.G. Kogan,
C.C. Tsuei and V. Kresin for useful discussions.

%\end{figure}

%\end{multicols}

\end{document}